\documentclass[12pt, a4paper]{article}
\usepackage{epsf}
\usepackage{cite}
\usepackage{amsmath,amssymb}
\input{colordvi.tex}
\usepackage[usenames,dvipsnames]{color}
\usepackage[dvips]{graphicx}
\usepackage{comment}
\begin{document}

\newcommand{\lsim}{\stackrel{<}{_\sim}}
\newcommand{\gsim}{\stackrel{>}{_\sim}}
\newcommand{\rem}[1]{{ {\color{red} [[$\spadesuit$ \bf #1 $\spadesuit$]]} }}

\renewcommand{\theequation}{\thesection.\arabic{equation}}
\renewcommand{\thefootnote}{\fnsymbol{footnote}}
\setcounter{footnote}{0}


\def\thefootnote{\fnsymbol{footnote}}
\def\a{\alpha}
\def\b{\beta}
\def\c{\varepsilon}
\def\d{\delta}
\def\e{\epsilon}
\def\f{\phi}
\def\g{\gamma}
\def\h{\theta}
\def\k{\kappa}
\def\l{\lambda}
\def\m{\mu}
\def\n{\nu}
\def\p{\psi}
\def\q{\partial}
\def\r{\rho}
\def\s{\sigma}
\def\t{\tau}
\def\u{\upsilon}
\def\v{\varphi}
\def\w{\omega}
\def\x{\xi}
\def\y{\eta}
\def\z{\zeta}
\def\D{\Delta}
\def\G{\Gamma}
\def\H{\Theta}
\def\L{\Lambda}
\def\F{\Phi}
\def\P{\Psi}
\def\S{\Sigma}
\def\me{\mathrm e}

\def\o{\over}
\def\beq{\begin{eqnarray}}
\def\eeq{\end{eqnarray}}
\newcommand{\vev}[1]{ \left\langle {#1} \right\rangle }
\newcommand{\bra}[1]{ \langle {#1} | }
\newcommand{\ket}[1]{ | {#1} \rangle }
\newcommand{\bs}[1]{ {\boldsymbol {#1}} }
\newcommand{\mc}[1]{ {\mathcal {#1}} }
\newcommand{\mb}[1]{ {\mathbb {#1}} }
\newcommand{\EV}{ {\rm eV} }
\newcommand{\KEV}{ {\rm keV} }
\newcommand{\MEV}{ {\rm MeV} }
\newcommand{\GEV}{ {\rm GeV} }
\newcommand{\TEV}{ {\rm TeV} }
\def\diag{\mathop{\rm diag}\nolimits}
\def\Spin{\mathop{\rm Spin}}
\def\SO{\mathop{\rm SO}}
\def\O{\mathop{\rm O}}
\def\SU{\mathop{\rm SU}}
\def\U{\mathop{\rm U}}
\def\Sp{\mathop{\rm Sp}}
\def\SL{\mathop{\rm SL}}
\def\tr{\mathop{\rm tr}}
\def\sp{\;\;}

\def\IJMP{Int.~J.~Mod.~Phys. }
\def\MPL{Mod.~Phys.~Lett. }
\def\NP{Nucl.~Phys. }
\def\PL{Phys.~Lett. }
\def\PR{Phys.~Rev. }
\def\PRL{Phys.~Rev.~Lett. }
\def\PTP{Prog.~Theor.~Phys. }
\def\ZP{Z.~Phys. }

\begin{titlepage}

\begin{center}

\hfill UT-15-42, TU-1014, IPMU16-0033\\

\vskip .75in

{\Large \bf 
Viable Chaotic Inflation as a Source of Neutrino \\
Masses and Leptogenesis
}

\vskip .75in

{\large Kazunori Nakayama$^{a,b}$, Fuminobu Takahashi$^{c,b}$ and Tsutomu T. Yanagida$^{b,d}$}

\vskip 0.25in

\begin{tabular}{ll}
$^{a}$&\!\! {\em Department of Physics, Faculty of Science, }\\
& {\em The University of Tokyo,  Bunkyo-ku, Tokyo 133-0033, Japan}\\[.3em]
$^{b}$ &\!\! {\em Kavli IPMU (WPI), UTIAS,}\\
&{\em The University of Tokyo,  Kashiwa, Chiba 277-8583, Japan}\\[.3em]
$^{c}$ &\!\! {\em Department of Physics, Tohoku University,}\\
&{\em Sendai, Miyagi 980-8578, Japan}\\[.3em]
$^{d}$ &\!\! {\em ARC Centre of Excellence for Particle Physics at the Terascale,}\\
&{\em School of Physics, The University of Melbourne, Victoria 3010, Australia}
\end{tabular}

\end{center}
\vskip .5in

\begin{abstract}
We show that the seesaw mechanism as well as leptogenesis are 
natural outcomes of a viable chaotic inflation in supergravity. 
The inflation model contains two superfields, the inflaton and stabilizer fields, which, being singlets under the standard model
gauge symmetry, naturally couple to the lepton and Higgs doublets. 
The inflaton decays into leptons and Higgs fields, and the reheating temperature is predicted to be of ${\cal O}(10^{13})$\,GeV,
for which thermal leptogenesis is possible. On the other hand, gravitinos are copiously produced, and various
solutions to the gravitino problem are discussed.
We also argue that, if the shift symmetry of the inflaton is explicitly broken down to a discrete one,
 neutrino Yukawa couplings are periodic in the inflaton field,
and masses of leptons and Higgs do not blow up even if the inflaton takes super-Planckian field values.
The inflaton potential is given by a sum of sinusoidal functions with different 
height and periodicity, the so-called multi-natural inflation.
We show that the predicted scalar spectral index and tensor-to-scalar ratio lie in the region favored by the Planck data.
\end{abstract}

\end{titlepage}


\renewcommand{\thepage}{\arabic{page}}
\setcounter{page}{1}
\renewcommand{\thefootnote}{\#\arabic{footnote}}
\setcounter{footnote}{0}


\section{Introduction}
\setcounter{equation}{0}

Cosmic microwave background (CMB) temperature and polarization anisotropies have coherence 
beyond the horizon at the last scattering. This clearly shows that our Universe has experienced
accelerated expansion, i.e.,  inflation,
at a very early stage of the evolution. In particular,  a single-field slow-roll inflation is consistent
with the observations. 

Among many inflation models so far, there is an interesting class of models 
called large-field or chaotic inflation~\cite{Linde:1983gd}.
The simplest chaotic inflation is based on the quadratic potential where
the inflaton mass is fixed to be order $10^{13}$\,GeV by the normalization of the curvature perturbations. 
One of the advantages of the chaotic inflation is that it has no initial condition problem.
With the chaotic initial condition at the Planckian epoch, some patch of the Universe 
will necessarily start to inflate. For this, the initial inflaton field value has to be larger than the Planck scale $M_P (\simeq
2.4 \times 10^{18}\,{\rm GeV})$
by many orders of magnitude, and it is customary to impose a shift symmetry on the inflaton 
to keep the inflaton potential under control.

After inflation ends, the inflaton must decay into the Standard Model (SM) particles to reheat the Universe.
Also,  a right amount of baryon asymmetry needs to be created after inflation because any pre-existing baryon
number is exponentially diluted by the inflationary expansion. These two issues are highly model-dependent, and
often treated separately from the inflation model building. 
 
 The purpose of the present letter is twofolds. First, we provide a viable chaotic inflation in supergravity,
which automatically explains neutrino masses and the origin of baryon asymmetry via leptogenesis~\cite{Fukugita:1986hr}. 
To illustrate the idea, let us consider a simple quadratic chaotic inflation in supergravity, which 
necessitates two singlet superfields~\cite{Kawasaki:2000yn}, 
the inflaton $(\Phi)$ and the stabilizer field $(S)$, with the superpotential
\beq
W_{\rm inf} = M S \Phi,
\eeq
where $M \sim 10^{13}$\,GeV is the inflaton mass. The K\"ahler potential is assumed to respect a shift symmetry of the
inflaton along the imaginary direction, 
\beq
\varphi \to \varphi + A,
\eeq
where $\varphi \equiv \sqrt{2} {\rm Im}[\Phi]$ and $A$ is the real transformation parameter.  The $\Phi$ and $S$
are singlets under the SM gauge symmetry, and therefore, they naturally 
couple to the lepton $(L)$ and Higgs $(H_u)$ doublets,
\beq
W_{\rm yukawa} \sim S L H_u + \Phi L H_u,
\eeq
where we have omitted coupling constants and the flavor indices. 
 Then, integrating out the heavy inflaton and stabilizer fields, one
can explain the light neutrino masses by the seesaw mechanism~\cite{seesaw}. Note that
the suggested inflaton mass of order $10^{13}$\,GeV is intriguingly close to the right-handed (RH) neutrino mass scale required by 
the seesaw mechanism. After inflation,
the inflaton  decays into leptons and Higgs fields, and the reheating temperature is 
considered to be so high that successful thermal leptogenesis is possible. 
Thus, the seesaw mechanism as well as leptogenesis are natural outcomes of the 
chaotic inflation.
As a result, both the inflaton and stabilizer fields can be identified with the
RH sneutrinos, and they are expressed as $N_1 \equiv \Phi$ and $N_2 \equiv S$ in the following.
 Note here that the neutrino oscillation data can be explained with only two 
RH neutrinos~\cite{Frampton:2002qc}. 

In this chaotic inflation with the seesaw mechanism, the inflaton as well as the stabilizer field have sizable neutrino Yukawa couplings of ${\cal O}(0.1)$,
and therefore, the chaotic initial condition with super-Planckian inflaton field values
cannot be realized. This is because the masses of the leptons and Higgs would 
exceed the Planck mass for the inflaton field value greater than ${\cal O}(10) M_P$, and then,
 the effective field theory description breaks down. 
Our second purpose is to propose a solution to the problem
  and study its implications.  In fact, the problem can be solved if
 the neutrino Yukawa couplings are some functions of the inflaton so that 
 the masses of the leptons and Higgs do not monotonically increase with the inflaton field;
 for example, they may asymptote to a constant value or start to decrease for super-Planckian inflaton field values.
 An interesting possibility is that the system comes back to the SM as the inflaton field exceeds a critical value.
 This is the case if the shift symmetry
 of the inflaton is not completely broken by the superpotential interactions, but there remains an unbroken discrete shift symmetry.
 That is to say, the superpotential interactions are invariant under the following discrete shift transformation along the imaginary direction,
\begin{align}
\varphi \to \varphi+2 \pi f,
\label{disc}
\end{align}
where $f$ is the decay constant.  In general, the inflaton potential is given by
a sum of sinusoidal functions with different height and potential, the so-called multi-natural 
inflation~\cite{Czerny:2014wza,Czerny:2014xja,Czerny:2014qqa}.\footnote{
A sizable running spectral index can be generated in large field inflation with 
modulations~\cite{Kobayashi:2010pz,Czerny:2014wua}. 
See also Refs.~\cite{Higaki:2014pja,Kappl:2015esy,Choi:2015aem}.
} 
To avoid the blow-up of the masses of leptons and Higgs, 
$f$ should be smaller than or comparable to ${\cal O}(10) M_P$.
Thus,  the prediction of the scalar spectral index $(n_s)$ as well as the tensor-to-scalar ratio $(r)$ are
 naturally deviated from the simple quadratic chaotic inflation. As we shall see, for $f = {\cal O}(10) M_P$, the predicted
 $(n_s, r)$ can lie in the range preferred by the Planck data.

Lastly let us briefly mention related works in the past. Phenomenological aspects of sneutrino chaotic inflation
was studied in Ref.~\cite{Ellis:2003sq} where the supergravity effects were neglected. 
A first attempt to build  a sneutrino chaotic inflation model in supergravity relied on a rather complicated
form of the K\"ahler potential~\cite{Murayama:1992ua,Nakayama:2013nya}. Another approach is based on
a Heisenberg symmetry~\cite{Binetruy:1987xj,Gaillard:1995az} (see  \cite{Antusch:2009ty}). 
A D-term hybrid inflation with sneutrino was studied in Ref.~\cite{Kadota:2005mt}.
More recently, a much simpler realization was proposed~\cite{Murayama:2014saa},
based on a generic construction of chaotic inflation models in supergravity~\cite{Kawasaki:2000yn} (see also \cite{Kallosh:2010ug}).
The recent Planck observations~\cite{Ade:2015lrj}, however, excluded the quadratic chaotic inflation model, 
which requires some modifications of the inflaton potential
such as the polynomial chaotic inflation~\cite{Nakayama:2013jka,Nakayama:2013txa,Nakayama:2014wpa}.
See also Refs.~\cite{Evans:2015mta} for this issue in a setup of the RH sneutrino chaotic inflation.

\section{Viable sneutrino chaotic inflation}
\setcounter{equation}{0}
The successful chaotic inflation in supergravity necessitates two singlet superfields, the inflaton and
stabilizer fields, which naturally couple to the lepton and Higgs doublets as they are singlets under
the SM gauge symmetry.  It implies that the seesaw mechanism is a built-in feature of the chaotic
inflation in supergravity. As a result the inflaton and the stabilizer field are identified with the RH
sneutrinos. In this section, we study in detail such sneutrino inflation model. In particular, we assume that
 the shift symmetry of the inflaton is broken down to its discrete one as Eq.~(\ref{disc}).

The K\"ahler and super-potentials relevant for the inflation model are
\begin{align}
	&K = \frac{1}{2}(N_1+N_1^\dagger)^2 + |N_2|^2 - k_{2}\frac{|N_2|^4}{M_P^2} + |L_\alpha|^2 + |H_u|^2,\\
	&W= M N_2\sum_{n=1}\frac{g_n}{2 n} f \left( e^{\frac{\sqrt{2}nN_1}{f}} -  e^{\frac{-\sqrt{2}nN_1}{f}} \right)
	+ \left[ y_{1\alpha} \sum_{n=1}\frac{g_n'}{2 n} f \left( e^{\frac{\sqrt{2} nN_1}{f}} -  e^{\frac{-\sqrt{2}nN_1}{f}} \right)
	+ y_{2\alpha}N_2 \right]  L_\alpha H_u,
	\label{Wfull}
\end{align}
where $\alpha$ runs over the lepton flavor $e,\mu$, and $\tau$, $k_2$ is a positive constant of order unity, and we take $g_1 = g_1'=1$.
Here and in what follows, the summation over repeated indices is understood. The above K\"ahler potential
respects a continuous shift symmetry along the imaginary direction of $N_1$, which ensures the flatness
of the inflaton potential at $\varphi \gtrsim M_P$. The shift symmetry is explicitly broken down to
a discrete one (\ref{disc}) by the above superpotential interactions. We also impose a
 $Z_2$ symmetry under which both $N_1$ and $N_2$ flip the sign.\footnote{
	This $Z_2$ can be identified with $Z_2^{\rm (B-L)}$ under which all quarks and leptons flip the sign.
}
Here we take a basis of $L_\alpha$ such that charged lepton yukawa sector is diagonalized
and omitted other interactions of the SUSY SM fields.
The RH neutrino mass parameter $M$ can be taken real and positive without loss of generality.
The coefficients $g_n$ and $g_n'$ are assumed to be suppressed for larger $n$.

For $|N_1|\ll f$, the model is reduced to
\begin{align}
	K  &= \frac{1}{2}(N_1+N_1^\dagger)^2 + |N_2|^2 - k_2\frac{|N_2|^4}{M_P^2},\\
	W &= M N_1 N_2 + y_{i\alpha} N_i L_\alpha H_u,
	\label{Wapp}
\end{align}
where $i=1,2$.
Since typical values of $y_{i\alpha}$ is of $\mathcal O(0.1)$ for reproducing the observed neutrino masses (see Sec.~\ref{sec:nu}), 
the masses of $L_\alpha H_u$ would exceed the Planck mass during inflation if this effective theory holds up to 
$|{\rm Im} N_1| \sim \mathcal O(10)M_P$.
However, thanks to the discrete shift symmetry, the actual superpotential is given by (\ref{Wfull}), where
the masses of $L_\alpha H_u$ are periodic with respect to  $\varphi (= \sqrt{2} {\rm Im} N_1)$.
Thus their masses remain smaller than the Planck mass, in which case 
we can safely discuss the inflaton dynamics.

\subsection{Inflaton potential}

In our model, the inflaton $\varphi$ is identified with the imaginary component of $N_1$,  
$\varphi = \sqrt{2}{\rm Im}[N_1]$. The inflaton potential is then given by
\begin{align}
	V = M^2 \left| \sum_n \frac{g_n}{n}\sin\left(\frac{n\varphi}{f}\right) \right|^2,
\end{align}
where we have taken ${\rm Re}[N_1]=N_2=H_u=L_\alpha=0$.
Since $N_2$ receives a Hubble mass of $m_{N_2}^2 = 12k_2 H^2$ with $H$ being the Hubble parameter 
and $\chi\equiv \sqrt{2}{\rm Re}[N_1]$ also receives $m_{\chi}^2 = 3H^2$ during inflation, they are heavy enough to be stabilized
around $N_2 = \chi = 0$.
To be precise, $\chi$ slightly shifts from zero, but its effect on the inflationary prediction is negligible~\cite{Nakayama:2014wpa}.
We will check the validity of the assumption of $H_u=L_\alpha=0$ later in Sec.~\ref{sec:stab}.
The inflaton potential is given by a sum of sinusoidal functions with
different height and periodicity, and it is the so-called multi-natural inflation~\cite{Czerny:2014wza}.

\begin{figure}
\begin{center}
\includegraphics[scale=1.2]{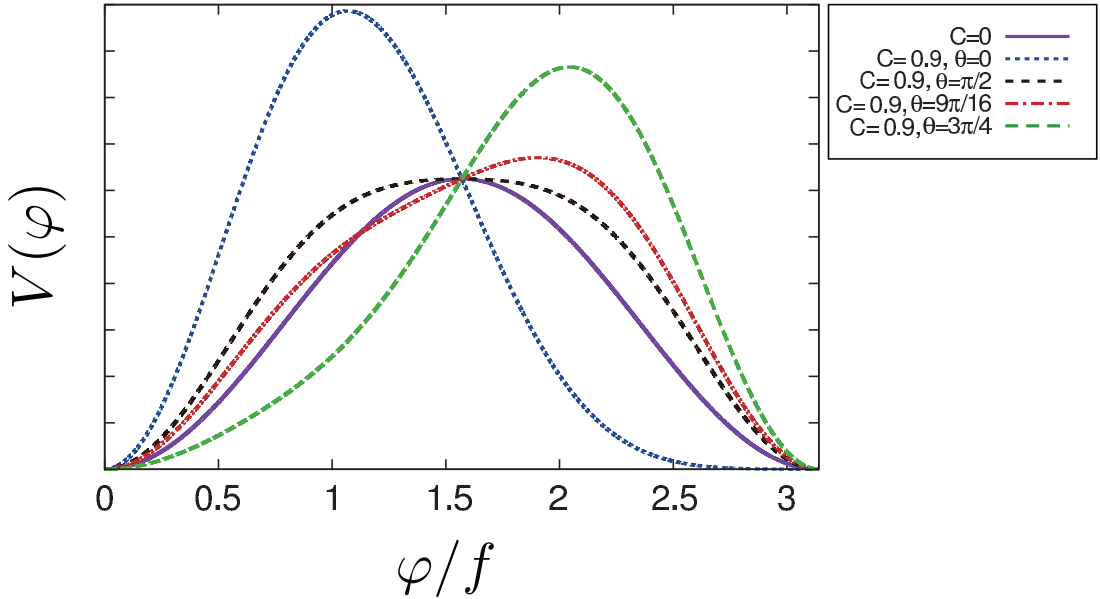}
\end{center}
\caption {
Shape of the scalar potential of $\varphi$ for  $C=0.9$ with $\theta = 0, \pi/2, 9\pi/16$, and $3\pi/4$ 
for fixed $f$ and $M$. The case of $C=0$ is also shown for comparison.
}
\label{fig:pot}
\end{figure}

To proceed, let us take the first two terms:
\begin{align}
	V &= \frac{1}{2}M^2 f^2 \left| \sin\left(\frac{\varphi}{f}\right)   +\frac{g_2}{2 }\sin\left(\frac{2\varphi}{f}\right) \right|^2,\nonumber\\
	&=	 \frac{1}{2}M^2 f^2\sin^2\left(\frac{\varphi}{f}\right)   \left[1
	+2C\cos\theta \cos\left(\frac{\varphi}{f}\right)
	+C^2 \cos^2\left(\frac{\varphi}{f}\right) 
	\right],
	\label{vphi}
\end{align}
where we have used $g_1 = 1$ and defined $g_2 \equiv Ce^{i\theta}$.
For $C=0$ this is nothing but a potential for natural inflation, but a nonzero  $C$ deforms the inflaton potential and 
the prediction of the spectral index $n_s$ and the tensor-to-scalar ratio $r$ are modified.
In Fig.~\ref{fig:pot} we show the shape of the scalar potential (\ref{vphi}) for  $C=0.9$ 
with $\theta = 0, \pi/2, 9\pi/16,$ and $3\pi/4$ for fixed $f$ and $M$. The case of $C=0$ is also shown for comparison.
We have numerically checked that there are no local minima of the potential in the field range $0< \varphi/f < \pi$
for $0\leq C < 1$.

\begin{figure}
\begin{center}
\includegraphics[scale=1.2]{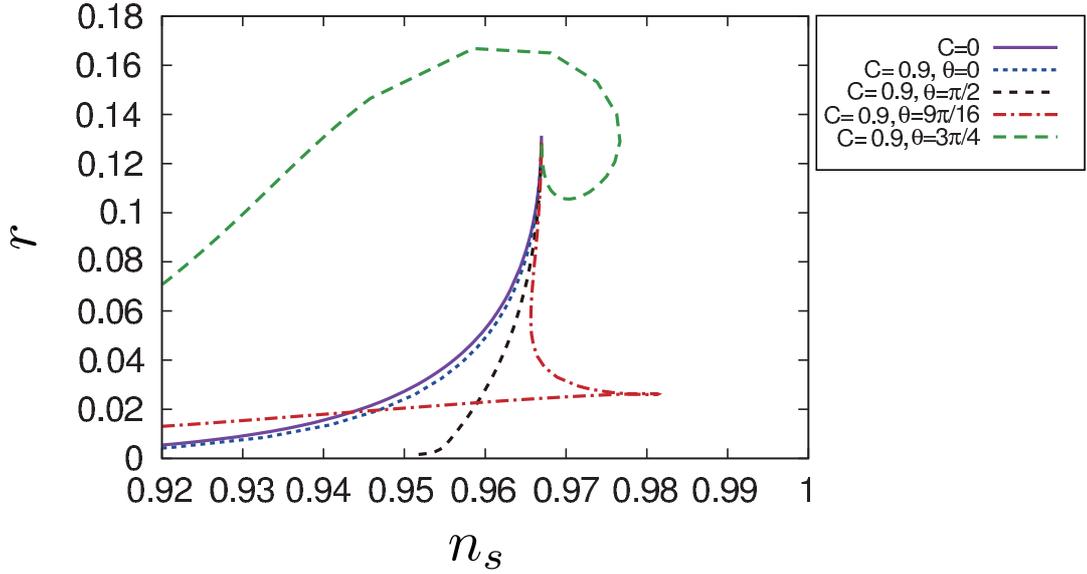}
\end{center}
\caption {
	The prediction of $(n_s,r)$ for the same parameters in Fig.~\ref{fig:pot}.
	For each line, we varie $f$ in the range $5M_P < f < 100 M_P$.
}
\label{fig:ns}
\end{figure}

We have numerically solved the slow-roll equation of motion
\begin{align}
	&\ddot\varphi + 3H \dot\varphi + V' = 0, \\
	& 3M_P^2 H^2 \simeq V(\varphi),
\end{align}
and calculated the slow-roll parameters,
\begin{align}
	\epsilon = \frac{M_P^2}{2}\left( \frac{V'}{V} \right)^2,~~~\eta = M_P^2\frac{V''}{V},
\end{align}
 at $60$ e-folding  before inflation ends.
Then the scalar spectral index and tensor-to-scalar ratio are given by
\begin{align}
	n_s = 1-6\epsilon + 2\eta,~~~r=16\epsilon.
\end{align}
The results are shown in Fig.~\ref{fig:ns} for the same parameters as in Fig.~\ref{fig:pot}.
For each line, we have varied $f$ in the range $5M_P < f < 100 M_P$. 
One can see that the predicted $n_s$ and $r$ significantly differ from those
of natural inflation, and that they can lie in the region favored by the Planck result,
$n_s=0.9655 \pm 0.058$ and $r < 0.09$~\cite{Ade:2015lrj}.
 The Planck normalization on the primordial curvature
perturbation can be satisfied by adjusting the overall scale of the inflaton potential,
which results in $M = {\cal O}(10^{13})$\,GeV unless $r$ is smaller than
${\cal O}(10^{-3})$.

\subsection{Stability of inflationary path}  \label{sec:stab}

In Ref.~\cite{Nakayama:2013txa} it was shown that the inflationary trajectory may be destabilized in
the presence of a coupling like $\lambda X H_u H_d$ with $X$ being a stabilizer field, 
because the $H_u H_d$ direction becomes tachyonic during inflation.
This constrains the coupling constant as $|\lambda| \lesssim 10^{-6}$.
In our case, the stabilizer field $N_2$ has large yukawa couplings to $L_\alpha H_u$
to reproduce the observed neutrino masses,
and hence one may think that this induces a similar instability.
Below we show that this instability does not exist thanks to the large yukawa couplings of $N_1$ to $L_\alpha H_u$.

The scalar potential of $L_\alpha$ and $H_u$ up to the quadratic terms during inflation is given by
\begin{align}
	V =\left[ MN_1^*\left(\sum_\alpha y_{2\alpha} L_\alpha\right)H_u + {\rm h.c.}\right]
	 +\left| \sum_\alpha y_{1\alpha} L_\alpha \right|^2 |N_1|^2
	 +\sum_\alpha \left| y_{1\alpha}\right|^2 |N_1|^2 |H_u|^2,
\end{align}
where we have taken $N_2=0$.
This is rewritten as
\begin{align}
	V =\left( M\overline{y_2}N_1^* L_2' H_u + {\rm h.c.}\right)
	 +\overline{y_1}^2 |N_1|^2 \left( \left| L_1' \right|^2 + |H_u|^2\right),
\end{align}
where
\begin{align}
	&\overline{y_1} \equiv \sqrt{ \sum_{\alpha} |y_{1\alpha}|^2 },~~~\overline{y_2} \equiv \sqrt{ \sum_{\alpha} |y_{2\alpha}|^2 }, \label{ybar}\\
	&L_1' \equiv \frac{1}{\overline{y_1} } \left( \sum_\alpha y_{1\alpha} L_\alpha  \right),
	~~~L_2' \equiv \frac{1}{\overline{y_2} } \left( \sum_\alpha y_{2\alpha} L_\alpha  \right).
\end{align}
Thus $L_1'$ obtains a mass of $\overline{y_1}|N_1|$ and hence is stabilized.
The mass matrix of $(H_u, L_2'^*)$ is given by
\begin{align}
	m^2_{H L_2'} = \begin{pmatrix}
			\overline{y_1}^2 |N_1|^2  & M \overline{y_2} N_1  \\
			 M \overline{y_2} N_1^*     &  k H^2
	\end{pmatrix},
\end{align}
where we have included a Hubble mass correction at the $(2,2)$ entry with $k$ being a positive constant of order unity.\footnote{
	This is easily achieved by introducing the K\"ahler potential $K = (1-k/3) |N_2|^2 |L_2'|^2 / M_P^2$.
	Other entries also receive Hubble mass corrections from similar terms, but they are subdominant since $|N_1| \gtrsim M_P$ during inflation.
}
This matrix has two eigenvalues of
\begin{align}
	M^2 \simeq \overline{y_1}^2 |N_1|^2,~~kH^2-\frac{\overline{y_2}^2}{\overline{y_1}^2} M^2,
\end{align}
for $\overline{y_1}|N_1| \gg \overline{y_2}M$.
Thus there are no tachyonic direction during inflation if $kH^2 > {\overline{y_2}^2M^2}/{\overline{y_1}^2}$.
Actually this condition is easily satisfied. 
Although the other combination of $L_\alpha$
orthogonal to both $L_1'$ and $L_2'$ remains massless at this level,
it can also have a positive Hubble mass by introducing $K = - |N_2|^2 |L_\alpha|^2/M_P^2$.
Therefore the inflationary path is stable and we can take $H_u = L_\alpha=0$ during inflation.

As we shall see in Sec.~\ref{sec:nu}, a typical value of $\overline y_1$ is of $\mathcal O(0.1)$
which may lead to super-Planckian masses for leptons and Higgs during inflation, spoiling
the effective field theory description.
This is our motivation to introduce a discrete shift symmetry on the inflaton field and we need $f \lesssim \mathcal O(10) M_P$.
We note, however, that it is in principle possible to have $\overline y_1 \lesssim \mathcal O(0.01)$.
In such a case, we can take $f = \mathcal O(100) M_P$ and the predicted $(n_s,r)$ are close to that of the quadratic
chaotic inflation.

\section{Implications} \label{sec:imp}
\setcounter{equation}{0}

\subsection{Neutrino masses and mixings} \label{sec:nu}

Here let us show that our model can reproduce the observed neutrino masses and mixings~\cite{Frampton:2002qc,Ibarra:2003up,Harigaya:2012bw}.
To this end, it is convenient to work with a basis in which the RH neutrino masses are diagonalized:
\begin{align}
	W = \frac{1}{2} \tilde M_i \tilde N_i \tilde N_i + \tilde y_{i \alpha} \tilde N_i L _\alpha H_u,  \label{WMdiag}
\end{align}
with
\begin{align}
	&\tilde M_1 = \tilde M_2 = M,\\
	&\tilde y_{1\alpha} = \frac{1}{\sqrt{2}}(y_{1\alpha} + y_{2\alpha}),~~~\tilde y_{2\alpha} = \frac{i}{\sqrt{2}}(-y_{1\alpha} + y_{2\alpha}),\\
	&\tilde N_1=\frac{1}{\sqrt{2}}( N_1+N_2),~~~\tilde N_2=\frac{i}{\sqrt 2}(N_1-N_2),
\end{align}
where we have expanded the interactions at the potential minimum.
After integrating out the RH neutrinos in \eqref{WMdiag}, we obtain
\begin{align}
	W = -\frac{1}{2} {\tilde y}_{i\alpha}{\tilde y}_{j\beta}(\tilde M^{-1})_{ij} (L_\alpha H_u)(L_\beta H_u).
\end{align}
Thus the light neutrino mass matrix is given by
\begin{align}
	m^{(\nu)}_{\alpha\beta}
	=\frac{v^2\sin^2\beta}{M} \tilde y_{i\alpha}\tilde y_{i\beta},
\end{align}
where $v =174\,$GeV and $\sin\beta\equiv \langle H_u\rangle / v$.
Note that since the mass matrix $(\tilde M^{-1})_{ij}$ is rank 2, $m^{(\nu)}_{\alpha\beta}$ can only have two non-zero eigenvalues.
Therefore, among the total $9$ real parameters to characterize the neutrino mass matrix (three masses, three mixing angles and three CP phases), one mass parameter and one phase vanish.
Thus we are left with $7$ low energy parameters.
On the other hand, ${\tilde y}_{i\alpha}$ is a general $2\times 3$ complex matrix having 12 real parameters,
but three phases can be absorbed by redefining the phase of $L_\alpha$.\footnote{
	The phase redefinition of $L_\alpha$ combined with those of right handed charged leptons $\overline E_\alpha$
	can leave charged lepton yukawa matrix real and diagonal.
}
Thus the total physical degrees of freedom in the neutrino sector
is $2$ (RH neutrino mass) $+(12-3)$ (neutrino yukawa: ${\tilde y}_{i\alpha}$) $=11$
 if the RH neutrino masses are taken freely.
Of these, the overall rescaling $\tilde M_i \to \lambda_i^2 \tilde M_i$ combined with ${\tilde y}_{i\alpha} \to \lambda_i {\tilde y}_{i\alpha}$
does not affect the light neutrino masses.
This rescaling has two parameters, hence we are left with the $9$ parameters in the high energy 
to parameterize the light neutrino masses and mixings.
This number is larger than the number of the  low energy parameters, which is $7$, hence 
a model with two RH neutrinos has enough parameters to reproduce the light neutrino mass matrix.

In our case, the RH neutrino mass matrix has only one parameter $M$ and its value is fixed by the normalization
of the primordial curvature perturbations. 
Since $M$ is fixed, there are  $9$ parameters in the neutrino yukawa sector, which cannot be reduced further 
by the rescaling. Hence the situation remains intact.

By using the MNS matrix~\cite{Maki:1962mu}, neutrino mass eigenvalues are expressed as
\begin{align}
	m_{\bar\alpha}^{(\nu)} \delta_{\bar{\alpha} {\bar \delta}}= U^{{\rm (MNS)}T}_{\bar\alpha \beta}  m^{(\nu)}_{\beta\gamma} U^{\rm (MNS)}_{\gamma \bar\delta},
\end{align}
where $\bar\alpha = 1,2,3$ denotes the mass eigenstate basis.
Here we impose $m_{1}^{(\nu)} <  m_{2}^{(\nu)}$, so that $m^{(\nu)}_1=0$ for the normal hierarchy (NH) and 
 $m^{(\nu)}_3=0$ for the inverted hierarchy (IH).
The MNS matrix is parametrized as
\begin{align}
	U^{\rm (MNS)}_{\alpha\bar\beta} = \begin{pmatrix}
		c_{12}c_{13}    &    s_{12}c_{13}   &  s_{13}e^{-i\delta} \\
		-s_{12}c_{23}-c_{12}s_{23}s_{13}e^{i\delta} &  c_{12}c_{23}-s_{12}s_{23}s_{13}e^{i\delta} & s_{23}c_{13} \\
		s_{12}s_{23}-c_{12}c_{23}s_{13}e^{i\delta}  & -c_{12}s_{23}-s_{12}c_{23}s_{13}e^{i\delta}  &c_{23}c_{13} 
	\end{pmatrix}
	\times{\rm diag}\left(1,e^{i\alpha/2},1\right),
\end{align}
where $c_{ij} = \cos\theta_{ij}$, $s_{ij} = \sin \theta_{ij}$, $\delta$ is the Dirac phase and $\alpha$ is the Majorana phase.
The neutrino yukawa is given by
\begin{align}
	  \tilde y_{i\alpha}\tilde y_{i\beta}
	 = \frac{M}{v^2\sin^2\beta} U^{{\rm (MNS)}*}_{\alpha \bar\gamma}  m^{(\nu)}_{\bar{\gamma}}  \delta_{\bar{\gamma } \bar{\delta}}
	U^{\rm (MNS)\dagger}_{\bar\delta \beta}.
	\label{yy}
\end{align}
As mentioned above, this does not uniquely determine all the matrix elements of $\tilde y_{i\alpha}$:
there are additional two degrees of freedom in the yukawa sector to determine the light neutrino mass matrix.
This can be explicitly seen by solving \eqref{yy} as~\cite{Casas:2001sr,Ibarra:2003up}
\begin{align}
	 \tilde y_{i\alpha} = \frac{M^{1/2}}{v \sin\beta} R_{i \bar\gamma} \sqrt{m_{\bar\gamma}^{(\nu)}} 
	  \delta_{\bar{\gamma } \bar{\delta}} U^{\rm (MNS)\dagger}_{\bar\delta \alpha},
\end{align}
where
\begin{align}
	R_{i\bar\gamma} = \begin{pmatrix}
		0 & \cos z & -\sin z \\
		0 & \sin z & \cos z
	\end{pmatrix}
	~~~{\rm for~~~NH},
\end{align}
and
\begin{align}
	R_{i\bar\gamma} = \begin{pmatrix}
		-\sin z & \cos z & 0 \\
		\cos z & \sin z  & 0
	\end{pmatrix}
	~~~{\rm for~~~IH},
\end{align}
with $z$ being an arbitrary complex parameter, corresponding to the additional degrees of freedom.

The best-fit values of the observed parameters are~\cite{Capozzi:2013csa}
\begin{align}
	&\Delta m_{12}^2 = 7.54\times 10^{-5}\,{\rm eV^2},~~~\Delta m_{23}^2 = 2.43\times 10^{-3}\,{\rm eV^2},\\
	&\sin^2 \theta_{12}= 0.308,~~~\sin^2 \theta_{23}= 0.437,~~~\sin^2 \theta_{13}= 2.34\times10^{-2}.
\end{align}
for NH, and
\begin{align}
	&\Delta m_{12}^2 = 7.54\times 10^{-5}\,{\rm eV^2},~~~\Delta m_{23}^2 = 2.38\times 10^{-3}\,{\rm eV^2},\\
	&\sin^2 \theta_{12}= 0.308,~~~\sin^2 \theta_{23}= 0.455,~~~\sin^2 \theta_{13}= 2.40\times10^{-2}.
\end{align}
for IH.
We can determine the yukawa matrix $y_{i\alpha}$ by using these values for arbitrary values of $\delta, \alpha$ and $z$.
Note that $\overline{y_1}$ and $\overline{y_2}$ in (\ref{ybar}) are independent of $\delta$ and $\alpha$.
Also they are independent of $z$ as long as $z$ is real and in such a case we have $\overline{y_1}=\overline{y_2}$. 
If $z$ is real, we obtain
\begin{align}
	&\overline{y_1}=\overline{y_2}=0.139~~~{\rm for~~NH},\\
	&\overline{y_1}=\overline{y_2}=0.180~~~{\rm for~~IH},
\end{align}
for $M=2\times 10^{13}$\,GeV and $\sin\beta=1$.
We regard them as ``typical'' values. On the other hand, if $z$ has an imaginary component, the prediction changes.
For example, for $z=i$ we obtain
\begin{align}
	&\overline{y_1}=0.0511,~~~\overline{y_2}=0.377~~~{\rm for~~NH},\\
	&\overline{y_1}=0.0660,~~~\overline{y_2}=0.488~~~{\rm for~~IH}.
\end{align}
By taking a large value of imaginary component of $z$, we can make a hierarchy between $\overline{y_1}$ and $\overline{y_2}$.
For $|z| \lesssim \mathcal O(1)$, $\overline{y_1}$ and $\overline{y_2}$ are of  $\mathcal O(0.01-1)$.

\subsection{Reheating}

Now let us consider the reheating after inflation.
In our model, the reheating process is slightly nontrivial because of the large yukawa coupling of the inflaton.
Just after inflation, the inflaton $\varphi$ begins a coherent oscillation with its amplitude of order $M_P$, 
and the coupled charged leptons and Higgs (and their superpartners) obtain masses of order $\sim y \varphi$
where $y$ collectively denotes the neutrino yukawa coupling.
Since this is much larger than the inflaton mass $M$, the perturbative decay of the inflaton is not kinematically allowed.
Instead, non-perturbative particle production, called preheating, happens when $\varphi$ passes the origin $\varphi \simeq 0$~\cite{Kofman:1994rk}.
The produced particles decay into lighter ones before the inflaton again moves back to $\varphi \simeq 0$~\cite{Felder:1998vq}.
For example, Higgs boson decays into quarks through yukawa couplings.

The effective decay rate of $\varphi$ through this process is estimated as~\cite{Mukaida:2012qn}
\begin{align}
	\Gamma_\phi \sim \frac{2y^2 M}{\pi^{7/2}g} \equiv b M,
\end{align}
where $g$ collectively denotes the Higgs coupling to lighter particles and $b\sim \mathcal O(0.01-0.1)$.
Thus after a few Hubble time after inflation, a significant fraction of the inflaton energy density is transferred to radiation:
\begin{align}
	\rho_r \sim b \rho_\phi.
\end{align}
If this process continues to produce radiation even after the radiation energy density begins to dominate the universe,
the reheating temperature is given by
\begin{align}
	T_{\rm R} \sim 10^{15}\,{\rm GeV} \, \left(\frac{b}{0.01}\right)^{1/2} \left( \frac{M}{2\times 10^{13}\,{\rm GeV}} \right)^{1/2}.
\end{align}

On the other hand this preheating process may become ineffective due to the thermal mass correction to the Higgs particles.
Then the main process that transfers the inflaton energy to radiation becomes the scattering of light particles in thermal plasma
with inflaton.
In our case, at the high inflaton amplitude regime $y\varphi \gtrsim T$, it is the effective inflaton coupling with 
SU(2) gauge bosons after integrating out heavy Higgs and leptons that is responsible for such an effective dissipation rate.
The oscillation-averaged dissipation rate is estimated as~\cite{Mukaida:2012qn}
\begin{align}
	\Gamma_\phi^{\rm (dis)} \sim \frac{b' y\alpha_W^2 T^2}{\tilde\varphi},	
\end{align}
where $b'$ is an order one numerical constant, $\alpha_W$ is the SU(2) fine structure constant and $\tilde\varphi$ denotes the
oscillation amplitude of $\varphi$.
The inflaton is thermalized when this rate becomes comparable to the Hubble scale.
This occurs at $H \sim (\alpha_W^2 b^{1/2} b' y)^{2/3} M \sim \mathcal O(10^{-3}) M$.
Thus the reheating temperature in this case is estimated to be
\begin{align}
	T_{\rm R} \sim \mathcal O(0.01)\times \sqrt{MM_P} \sim 10^{14}\,{\rm GeV}\left( \frac{M}{2\times 10^{13}\,{\rm GeV}} \right)^{1/2}.
\end{align}
In either case, the reheating temperature is so high that the RH sneutrino inflaton becomes thermalized.

\subsection{Leptogenesis}

Now let us consider implication for the leptogenesis scenario~\cite{Fukugita:1986hr}.
In our model, two RH neutrino masses are nearly degenerate, and so, we must take account of the resonant effect~\cite{Covi:1996fm,Pilaftsis:1997dr,Buchmuller:1997yu,Pilaftsis:2003gt,Anisimov:2005hr,Garny:2011hg}.
It should be noticed that the effective CP asymmetry in the RH neutrino decay vanishes in the exact degenerate limit.\footnote{
	This can be seen in \eqref{Wapp} that we can rotate a phase of $N_1$ and $N_2$ without affecting the RH mass term,
	while it can absorb the phase of yukawa matrix which appears in the CP asymmetric decay of RH neutrinos.
	This phase rotation is allowed only for the degenerate case, i.e., there are no diagonal elements in the mass matrix of RH neutrinos.
}
Thus we need a small diagonal mass matrix element of $\sim \delta M$ $( |\delta M|\ll M)$ in (\ref{Wapp}).\footnote{
	As long as $|\delta M| \ll 10^{-2} M$, it does not much affect the inflaton dynamics.
}
After diagonalizing the RH neutrino mass matrix, we obtain
\begin{align}
	&W = \frac{1}{2}\tilde M_i \tilde N_i \tilde N_i + \tilde y_{i\alpha} \tilde N_i L_\alpha H_u,\\
	&\tilde M_1=M-\delta M,~~~\tilde M_2=M+\delta M.
\end{align}
The lepton asymmetry is generated via the CP asymmetric decay of $\tilde N_1$ and $\tilde N_2$.
The CP asymmetry parameter is given by
\begin{align}
	\epsilon_i = \frac{{\rm Im}\left[ (\tilde y_{i\alpha} \tilde y^\dagger_{\alpha j} )^2\right]}{8\pi  (\tilde y_{i\alpha}\tilde y^\dagger_{\alpha i} )}
	\frac{\tilde M_i\tilde M_j(\tilde M_i^2- \tilde M_j^2)}{(\tilde M_i^2- \tilde M_j^2)^2 +R^2},
\end{align}
where $R$ denotes the regulator, which is of the order of $\sim M \Gamma$ with
$\Gamma$ being the decay width of RH neutrino~\cite{Garny:2011hg,Garbrecht:2011aw,Iso:2013lba,Garbrecht:2014aga}.
The $\epsilon_i$ parameter is maximized and can be $\mathcal O(1)$ for $\tilde M_1^2- \tilde M_2^2 \sim R$.

The final baryon asymmetry, after the sphaleron conversion of the lepton number to the baryon number, is given by
\begin{align}
	\frac{n_B}{s} = \frac{8}{23}\frac{n_L}{s} = \frac{8}{23}\kappa \sum_{i=1,2} \epsilon_i\frac{n_{N_i}}{s},
\end{align}
where $\kappa$ represents the suppression factor due to the washout effect~\cite{Giudice:2003jh}.
In our case, RH neutrinos are expected to be in thermal equilibrium at $T\sim M_i$:
$\Gamma_{N_1}/H_{T=M_i} \sim \mathcal O(100)$. Thus we have $\kappa \sim \mathcal O(0.01)$.
To explain the observed value $n_B/s \simeq 9\times 10^{-11}$, we need $\epsilon_i\sim 10^{-5}$.\footnote{
	The calculation based on the Boltzmann equation may become invalid for the strongly degenerate case. 
	Instead, we may need Kadanoff-Baym approach to estimate the lepton asymmetry.
	In any case, we can obtain small $\epsilon_i$ in a degenerate limit~\cite{Garny:2011hg}.
}

\subsection{Gravitino problem}

Finally we discuss the gravitino problem.
In general, there are two contributions to the gravitino production: thermal production and nonthermal production.
The nonthermal gravitino production rate from the direct decay of the inflaton depends on the inflaton vacuum expectation value~\cite{Kawasaki:2006gs,Asaka:2006bv,Endo:2006tf,Endo:2007ih,Endo:2007sz}.
In our model, thanks to the $Z_2$ symmetry, the inflaton VEV is zero and hence there is no significant production of the gravitino
from the inflaton decay.

Thermal gravitino production, on the other hand, is very efficient in the present model because of high reheating temperature.
The gravitino abundance in terms of the number to the entropy density ratio is estimated as~\cite{Bolz:2000fu}
\begin{align}
	\frac{n_{3/2}}{s} \simeq 2\times 10^{-12}\left(1+\frac{m_{\tilde g}^2}{3m_{3/2}^2} \right)\left( \frac{T_{\rm R}}{10^{10}\,{\rm GeV}} \right),
\end{align}
where $m_{\tilde g}$ and $m_{3/2}$ denote the gluino mass and gravitino mass, respectively.
The cosmological consequences depend on the gravitino mass.
For the unstable gravitino, in order to avoid the constraint from big-bang nucleosynthesis (BBN)~\cite{Kawasaki:2004qu},
we need $m_{3/2}\sim 100$\,TeV so that it decays before BBN begins.
However, even in such a case, the lightest SUSY particles (LSP) produced by the gravitino decay give a too large contribution 
to the relic dark matter abundance independently of the gravitino mass, 
if we assume the anomaly mediation relation between the gaugino and gravitino mass~\cite{Randall:1998uk}.

A solution to the LSP overproduction is to introduce a small R-parity violation so that LSP decays quickly before BBN~\cite{Barbier:2004ez}.
Another way to avoid the gravitino problem is to assume an ultra light gravitino scenario in which $m_{3/2}\lesssim 4.7$\,eV~\cite{Pierpaoli:1997im,Viel:2005qj,Osato:2016ixc}.\footnote{
	For a model to obtain 125\,GeV Higgs boson mass for such light gravitino scenario, see e.g. Refs.~\cite{Ibe:2010jb,Yanagida:2012ef}.
}
A small amount of late-time entropy production after the LSP freezeout also helps the situation.
In such a case, baryon asymmetry is also diluted but in the present scenario it is easy to create larger amount of lepton asymmetry
due to the resonant effect.

\section{Conclusions}
\setcounter{equation}{0}
In this Letter we have shown that the seesaw mechanism as well as thermal leptogenesis are built-in features
of the chaotic inflation in supergravity. A successful chaotic inflation in supergravity requires two superfields, the inflaton
and the stabilizer fields, which, being singlets under the SM gauge symmetry, naturally couple to the lepton and Higgs
doublets. The typical mass scale of the inflaton and stabilizer fields is of order $10^{13}$\,GeV, which is fixed by the
normalization of the curvature perturbations. Integrating out the heavy inflaton and stabilizer fields, then, one naturally
explains the light neutrino masses via the seesaw mechanism. Also, the inflaton decays into leptons and Higgs after
inflation, and the reheating temperature is considered to be so high that thermal leptogenesis works. 
Thus, the inflaton and stabilizer fields are subsequently identified with the RH sneutrinos.

There is one potential problem of the sneutrino chaotic inflation model. As the inflaton has initially super-Planckian
field values, the effective masses of the lepton and Higgs fields may exceed the Planck mass, and the effective field
theory description may break down. To avoid the super-Planckian masses of the leptons and Higgs during inflation, 
we have assumed that the shift symmetry of the inflaton is not completely broken by the superpotential interactions,
but there remain an unbroken discrete shift symmetry, which modifies the shape of the inflaton potential.
As a result, we can obtain the prediction of $(n_s,r)$ within $1\sigma$ range of the Planck result if the decay
constant $f$ is ${\cal O}(10)M_P$.
While the reheating temperature is predicted to be so high that leptogenesis successfully works,
 gravitinos are also copiously produced, which leads to a cosmological problem.
Several solutions to the gravitino problem were also discussed. 

Since we have only two RH neutrinos, one of the light neutrinos is massless.
Therefore, $\langle m_{ee} \rangle = \left| \sum_{\bar \alpha} m_{\bar\alpha} U_{e\bar\alpha}^2 \right|$,
which is a quantity directly probed by the neutrinoless double beta decay experiments,
is bounded below in our model and predicted to be $\langle m_{ee} \rangle \sim 1\,$meV for NH and 
$\langle m_{ee} \rangle \sim 10\,$meV for IH with a slight dependence on the Majorana phase $\alpha$.
Although the rate of neutrinoless double beta decay is  suppressed compared to
the (quasi) degenerate case and below the current sensitivity ($\sim 0.1$\,eV), 
it is within the reach of forthcoming experiments ~\cite{GomezCadenas:2010gs,Barea:2013bz,Benato:2015via}.
Inflaton has rather large yukawa couplings to explain observed neutrino masses and hence 
the reheating automatically happens.
The predicted reheating temperature is high and leptogenesis works efficiently.

\section*{Acknowledgments}

T. T. Y. thanks Prof. Raymond Volkas for the hospitality during his stay at the University of Melbourne.
This work was supported by the Grant-in-Aid for Scientific Research on Scientific Research A (No.26247042 [KN and FT]),
Scientific Research B (No. 26287039 [FT and TTY]),
Young Scientists B (No.26800121 [KN], No. 24740135 [FT]) and Innovative Areas (No.26104009 [KN], No.15H05888 [KN], No. 23104008 [FT],
and No.15H05889 [FT]).
This work was supported by World Premier International Research Center Initiative (WPI Initiative), MEXT, Japan.



\end{document}